\documentclass{ws-p9-75x6-50}

\def\HO{{\rm H_0}}

\def\kmsmpc{{\rm km \, sec^{-1} \, Mpc^{-1}}}
\def\deg2{{\rm deg^2}}
\def\K{{\, \rm K}}

\def\cm2{\,{\rm cm}^{-2} }

\def\hkpc{\,h^{-1}\,{\rm kpc} }

\def\ergcms{\,{\rm erg\, cm^{-2}\, s^{-1}} }

\def\ergs{\,{\rm erg\, s^{-1}}}
\def\ergsh2{\,{\rm erg\, s^{-1}}, h=\frac{1}{2} }

\def\Jy{\,{\rm Jy} }

\def\muK{\,\mu{\rm K} }

\def\eg{{\it e.g., }}

\def\etal{{\it et al. }}

\begin{document}

%\title{An Improved Measurement of the Hubble Constant from the Sunyaev-Zeldovich Effect}
\title{The Hubble Constant from SZE Measurements in Low-Redshift Clusters}

\author{Brian S. Mason, Steven T. Myers, \& A.C.S. Readhead}

\address{105-24 Caltech \\
Pasadena, CA 91125 \\
Email:  {\tt bsm@astro.caltech.edu}}

\maketitle

\abstracts{ We present a determination of the Hubble constant from
measurements of the Sunyaev-Zeldovich Effect (SZE) in an
orientation-unbiased sample of 7 $z < 0.1$ galaxy clusters.  With
improved X-ray models and a more accurate 32-GHz calibration, we
obtain $\HO = 64^{+14}_{-11}\pm 14_{sys} \kmsmpc$ for a standard CDM
cosmology, or $\HO = 66^{+14}_{-11}\pm 15_{sys} \kmsmpc$ for a flat
$\Lambda$CDM cosmology.  The random uncertainty in this result is
dominated by intrinsic CMB anisotropies, and the systematic
uncertainty is chiefly due to uncertainties in the X-ray spectral
models.  The implications of this result and future prospects for the
method are discussed.
}

\section{Introduction}
For over two decades it has been known that the combination of X-ray
and SZE observations of rich galaxy clusters, under the assumption of
spherical symmetry, yields a direct measurement of the cosmic distance
scale\cite{Silk_and_White_1978,Cavaliere_et_al_1979}.  This 
comes about due to the fact that the X-ray brightness of a cluster
depends upon a different power of $h$ than the SZE (which is
unaffected by cosmological surface brightness dimming), allowing the
radio decrement, $\Delta T_{obs} \equiv q$, to be predicted from the
X-ray data up to a power of $h$: $\Delta T_{pred} \equiv p \, h^{-1/2}$.
Then $h$ is given by
\begin{equation}
\label{eq:pq}
h = \left( \frac{p}{q} \right)^2.
\end{equation}
Only in the past few years have reliable and accurate applications of
this method become possible
\cite{Birkinshaw_et_al_1991,Herbig_et_al_1995,Carlstrom_Joy_and_Grego_1996,Grainge_et_al_1996,Holzapfel_et_al_1997}.
Due to the assumption of spherical symmetry in this analysis,
selection biases have been a great concern in using the SZE as a
distance measure.  To address this concern Myers {\it et al.} (1997)
defined an X-ray flux-limited sample of 11 $z < 0.1$ clusters and
began a campaign to measure distances to these clusters.  With this
strategy, random departures from spherical symmetry in individual
clusters will not bias the sample average for ${\rm H_0}$.  With
observations of 4 clusters Myers et al. obtain a Hubble Constant of
$54 \pm 14 \, \kmsmpc$.  The accuracy of this result is limited by a
7\% radio calibration uncertainty and estimated 15-30\% X-ray model
uncertainties for each cluster.  In this proceeding, we present more
recent observations in the campaign initiated by Myers {\it et al.},
and discuss other improvements in these results.

Our efforts are concentrated on reducing and understanding the
uncertainties in the X-ray models, since these have always been a
limiting consideration for the SZE.  This is facilitated by the advent
of abundant X-ray data from ROSAT and other satellites.  In addition,
we have conducted an absolute radio calibration program at the Owens
Valley Radio Observatory (OVRO) which has reduced our calibration
uncertainties by over a factor of two.  We have also quantitatively
evaluated the effect of CMB anisotropies on our observations.

Throughout this proceeding we we use ${\rm H_0} = 100 \, h^{-1}
\kmsmpc$.  We will consider two cosmologies: SCDM ($\Omega_m=1,
\Omega_{\Lambda}=0$), and $\Lambda {\rm CDM}$ ($\Omega_m=0.3,
\Omega_{\Lambda}=0.7$).

\section{Cluster Sample}

In defining a sample for SZE-based $\HO$ determinations the most
important consideration is that the sample be free of any orientation
bias, such as would be introduced,\eg , by a surface brightness
selection.  This is especially important due to the fact
(Eq~\ref{eq:pq}) that systematic errors in the observations enter
quadratically into the final $\HO$ result.  Since X-ray cluster
catalogs are most reliable at high flux levels--- and the optical
catalogs which underlie most current X-ray cluster catalogs are most
reliable at low redshifts--- a low-$z$ sample is indicated.  This has
the additional advantage that the X-ray models will be better since,
at low-$z$, the signal is stronger and angular resolution is less of
an issue.

Mason \& Myers (2000) define an expanded, X-ray flux-limited cluster
sample of $z<0.1$ objects.  This is a 90\% volume-complete sample
selected from the XBAC catalog \cite{Ebeling_et_al_1996} with the
restrictions $F_X > 1.0 \times 10^{-11} \ergcms$.  At $z=0.1$, the
volume-completeness criterion corresponds to $L_X > 1.13 \times
10^{44} h^{-2} \ergs$ (0.1 - 2.4 keV).  The resulting set of 31
clusters contains the Myers \etal sample.  Although the XBAC sample is
ultimately derived from an optical catalog, the bright, rich clusters
which pass our luminosity criterion should be reliably selected in any
orientation.  The 7 clusters for which we present measurements here
are all members of the smaller Myers {\it et al.} sample, but this
work is part of an ongoing project to survey distances to the objects
in our expanded sample.

\section{ X-ray Models}

\begin{figure}[t]
\begin{center}
% will enlarge or reduce the postscript figures based on the xsize
\epsfxsize=3in 
% postscript image file name
%\epsfbox{scatterplot_small.ps} 
\leavevmode
\epsfbox{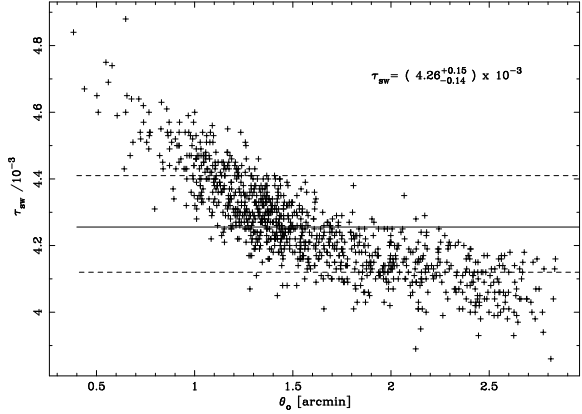}
\end{center}
\caption{Predictions of the 5.5-meter beam-convolved SZE decrement for
A2142.  The solid line shows the distribution mean and the dashed
lines show two-sided 68\% confidence regions.\label{fig:szscatter}}
\end{figure}

Mason \& Myers (2000) also present X-ray models for the 22 of the 31
clusters which had public ROSAT PSPC data as of May 1999.  The primary
focus of this analysis was to quantify the uncertainty in the
beam-convolved inverse Compton optical depth, $\tau_{sw}$.  To do
this, the ROSAT PSPC data were smoothed, and $10^3$ realizations of
each cluster were generated including Poisson noise; each realization
was analyzed just as we analyzed the real data.  We used a similar
strategy to quantify errors induced by inaccurate vignetting
correction and background subtraction.  We find, overall, that the
beam-convolved inverse-Compton optical depths are robustly constrained
by the ROSAT data, and unaffected by realistic PSPC systematics.  As
an example, Figure~\ref{fig:szscatter} shows the $10^3$ model
predictions for the beam-convolved optical depths ($\tau_{sw}$) from
our analysis of the A2142 data.  This is an extreme example since we
have excised the central region of the PSPC data, which is
contaminated by cooling flow emission, and allowed the fit to be
completely unconstrained there.  The result is that the core radius
$\theta_o$ is poorly determined; nevertheless the overall optical
depth is well constrained since the SZ decrement is mostly determined
by the extended lines of sight.  The cancellation of the large
uncertainties in the individual parameters $(\beta,\theta_o,n_{eo})$ comes
about due to strong parameter correlations, and it is generally
important to account for these ({\it e.g.}, by Monte Carlo methods) in
order to understand the uncertainties in the observables.

A summary of the models for the 7 clusters on which we have radio
observations is presented in Table~\ref{tbl:xmodels}.  For Coma,
following Hughes, Gorenstein, \& Fabricant (1988), we adopt a hybrid
temperature profile which is isothermal inside of $500 \hkpc$ and
adiabatic ($\gamma = 1.5$) outside of this.  For the other clusters we
assume an isothermal profile.  We use the global, cooling-flow
corrected electron temperatures of Markevitch \etal (1998).  Our value
for $\HO$ is not sensitive to the temperature profile we assume: it
changes by $< 5\%$ if we assume hybrid profiles for all of the
clusters.  While more extreme temperature profile models would have a
stronger affect on our result, such models are not motivated by
current analyses\cite{Irwin_Bregman_and_Evrard,White_2000}.

\begin{table*}[t]
\begin{center}
\caption{Cluster Model Parameters}
\label{tbl:xmodels}
\begin{tabular}{ l l l l l l }
\hline\hline
Cluster &  $z$ & $\theta_o$ & $\beta$ & $n_{eo}$ & $k T_{e}$  \\ 
        &      &  ($\,'\,$) &         & ($10^{-3} \, h^{1/2} \, {\rm
cm^{-3}}$) & (keV) \\ \hline
A399  &  $0.0715$  & $4.33 \pm 0.45$     & $0.742 \pm 0.042$      &
$3.23 \pm 0.18$  & $7.0 \pm 0.2$  \\
A401  & $0.0748 $    &  $2.26 \pm0.41$  &  $0.636 \pm 0.047 $    &
$7.90 \pm 0.81$ &  $8.0 \pm 0.2$  \\
A478  & $0.0900$ & $1.00 \pm 0.15$ & $0.638 \pm 0.014$ & $27.81 \pm
9.7$  &  $8.4 \pm 0.7 $  \\
A1651 &  $0.0825$ & $2.16 \pm 0.36$ & $0.712 \pm 0.036$ & $7.14 \pm
3.20$  &  $6.1 \pm 0.2 $  \\
Coma  & $0.0232$ & $9.32 \pm 0.10$  &  $0.670 \pm 0.003$ &  $4.52 \pm
0.04$ & $9.1 \pm 0.4$ \\
A2142 & $0.0899$ &   $1.60 \pm 0.12$ & $0.635 \pm 0.012$ &  $14.95 \pm
1.0$ &  $9.7 \pm 0.8$ \\
A2256 & $0.0601$  &  $5.49 \pm 0.21$ &  $0.847 \pm 0.024$ & $4.08 \pm
0.08$  &  $6.6 \pm 0.2$    \\ \hline
\end{tabular}
\end{center}
\end{table*}

\section{Radio Observations}

\subsection{Recalibration}

In March through May of 1998 we undertook a program of absolute
calibration with the OVRO 1.5-meter radio telescope.  The gain of this
antenna was measured by NIST's Boulder, CO near-field test range in
May of 1990; these measurements in conjunction with accurate total
power determinations on-site at OVRO allowed a direct determination of
the flux densities of celestial sources.  The primary target for this
program was the supernova remnant Cassiopeia A (Cas A), the second
brightest extrasolar source in the sky at cm wavelengths.  We find an
epoch 1998 flux density $S_{Cas A,1998} = 194 \pm 4 \Jy$ at 32.0 GHz.
Observations of Jupiter relative to Cas A with the OVRO 5.5-meter
telescope give $T_{Jup} = 152 \pm 5 \K$ at 32.0 GHz.  All of our SZ
measurements are referred to this flux density scale, which improves
by a factor of two over the uncertainty present in the Myers \etal
calibration.  The absolute calibration measurements are discussed at
length in Mason \etal (1999), and are the basis of the flux density
scales which have been adopted by
Saskatoon\cite{Sask,Miller_et_al_1999}, the 30 GHz channel of
MAT\cite{Miller_et_al_1999}, and the CBI\cite{Padin_et_al_2000}; the
RING5M\cite{Leitch_et_al_2000} calibration is comparable but
independent.

\subsection{SZE Observations \& Effect of the Intrinsic Anisotropy}

From October 1996 through March 1998 we observed A399, A401, and A1651
with the OVRO 5.5-meter telescope.  The results of these observations,
together with measurements of Coma\cite{Herbig_et_al_1995} and A2142, A2256, and
A478\cite{Myers_et_al_1997}, are shown in Table~\ref{tbl:observations}.  The
errors reported in this table are the observational noise only.  The
signficance of the detection $\sigma$ is also reported, including the
effects of intrinsic CMB anisotropies.

Intrinsic anisotropies on this scale ($\ell = 590 \pm 200$) are
known to be significant, although these will be averaged down somewhat
by our scan pattern.  To evaluate the impact of these anisotropies, we
generated $10^3$ realizations of $4 \, \deg2$ patches of sky from a
$\Lambda$CDM power spectrum normalized to the
RING5M\cite{Leitch_et_al_2000} observed power level.  Each realization
was convolved with the 5-m main beam and the switching pattern
characteristic of a typical SZE observations.  We find a residual CMB
signal of $\sigma_{cmb} = 64 \, \muK$, or $\sigma_{cmb} = 1.24 \times
10^{-5}$ in terms of the Compton-y, which is accounted for in our
$\HO$ analysis.

\begin{table}[t]
\begin{center}
\caption{OVRO SZE Observations}
\label{tbl:observations}
\begin{tabular}{l l l}
\hline\hline
Cluster & $y_{obs}$ & significance \\ 
        & $(10^{-5})$ & ($\sigma$) \\ \hline
A399    & $3.24 \pm 0.41$  & 2.5 \\
A401    & $6.93 \pm 0.46$  & 5.2 \\
A1651   & $4.88 \pm 0.59$   & 5.6 \\
Coma    & $6.38 \pm 1.01$   & 4.0 \\
A2142   & $9.10\pm0.52$     & 6.7 \\
A2256   & $5.00 \pm 0.60$    & 3.6 \\ \hline
\end{tabular}
\end{center}
\end{table}

\section{Interpretation \& Discussion}

When the effects of intrinsic CMB anisotropy are included, faint
clusters like A399 are seen to be detected at less than $3\sigma$.  It
is important to keep such clusters in the analysis, however, in order
that an orientation bias not be introduced into our result.  To
account for these low signal-to-noise measurements, we have developed
a maximum likelihood method of estimating $\HO$ from the SZ and X-ray
data.  Preliminary results from the application of this method to our
7 clusters are shown in Table~\ref{tbl:hores}.  The calibration
uncertainties are $3\%$ (radio) and $8\%$ (X-ray), and we estimate a
$10\%$ uncertainty in the SZE predictions due to the possibility of
substructure and non-isothermality in the ICM.  We find $\HO =
64^{+14}_{-11}\pm 14_{sys} \kmsmpc$ (SCDM), or $\HO =
66^{+14}_{-11}\pm 15_{sys} \kmsmpc$ ($\Lambda$CDM).  Final results
on this sample are in press\cite{mason_sz}.

\begin{table*}[b]
\centering
\caption{${\rm H_0}$ results on 5 Clusters}
\begin{tabular}{l l l c}
\hline\hline
Cluster & $y_{obs} $ & $y_{pred} $  & ${\rm
H_0}$ \\ 
 &$(10^{-5})$ & $(10^{-5} h^{-1/2})$ & $(\kmsmpc)$ \\ \hline
A399 & $3.24 \pm 0.41$ & $3.27 \pm 0.20$      & $102^{+116}_{-53}$ \\
A401 & $6.93 \pm 0.46$ & $4.82 \pm 0.32$      & $48^{+28}_{-16} $ \\
A478 & $7.77 \pm 0.58$ & $6.05 \pm 0.54$      & $61^{+33}_{-20}$ \\
A1651 & $4.88 \pm 0.59$ & $2.92 \pm 0.17$     & $36^{+35}_{-15} $ \\
Coma & $6.38 \pm 1.01$ & $5.00 \pm 0.38$     & $62^{+49}_{-24}$ \\
A2142 & $9.10\pm0.52$   & $8.12  \pm  0.74 $  & $79^{+34}_{-24}$ \\
A2256 & $5.00 \pm 0.60$ & $ 4.11  \pm  0.26$  & $67^{+62}_{-28}$ \\ 
SAMPLE & --- & ---  & ${\bf 64^{+14}_{-11}}$ \\ \hline
\multicolumn{4}{l}{NOTE: Observational and model ncertainties are $1\sigma$ random errors only.}\\
\end{tabular}
\label{tbl:hores}
\end{table*}

The ages implied [$(10.2 \pm 3.2) \times 10^9\,$ years SCDM; $(14.2
\pm 4.5)\times 10^9 \, $ years ${\rm \Lambda CDM}$] are consistent
with recent age determinations from main-sequence fitting, which give
ages of $(> 12 \pm 1)\times 10^9 \,$ years
\cite{Reid_1997,Chaboyer_et_al_1998}.  From a hydrostatic analysis of
22 X-ray clusters in our sample\cite{Mason_and_Myers_2000}, together
with this value for $\HO$, we determine $\Omega_M = 0.33 \pm 0.05$.
This argues strongly against SCDM cosmologies, and in light of recent
CMB measurements\cite{Miller_et_al_1999,Leitch_et_al_2000,boom} which
indicate $\Omega_{tot} \sim 1$, may suggest another form of energy
density such as the cosmological constant or
quintessence\cite{Caldwell_Dave_and_Steinhardt_1998}.  These results
are broadly in agreement with the concordance
models\cite{Bahcall_Ostriker_and_Steinhardt_1999} suggested by a
number of independent means.

Coverage of clusters in this sample is continuing with the Cosmic
Background Imager\cite{Udomprasert,cbiMG9} (CBI).  The CBI's greater
coverage ($\sim 20$ clusters) will significantly reduce uncertainties
due to intrinsic CMB anisotropies.  Although the results we report
here incorporate improved X-ray models, the modelling uncertainties
will still be the limiting consideration for larger samples.  In order
to achieve a 10\% determination of ${\rm H_0}$ from the SZE, a better
understanding of the spectral models for the cluster atmospheres will
then be essential.

\section*{Acknowledgments}

We are grateful to Russ Keeney for his work on the OVRO 5-meter and
40-meter telescopes, to Jon Sievers for help with the 1997/98 OVRO
observations and the CMB simulations, and to Erik Leitch for writing
the OVRO data-analysis software.  We received suport from NSF grants
AST 91-19847 and AST 94-19279.  STM was supported by an Alfred
R. Sloan Fellowship at the University of Pennsylvania.

\end{document}